\documentclass[twoside]{article}

\usepackage[accepted]{aistats2022}
\usepackage[dvipsnames]{xcolor}
\usepackage{graphicx}
\usepackage{subfig}
\usepackage[round]{natbib}

\usepackage{aas_macros}

\begin{document}

%

%

\twocolumn[
    
\aistatstitle{An Unsupervised Hunt for Gravitational Lenses}

\aistatsauthor{ Stephen Sheng \And Keerthi Vasan G.C. \And Chi Po Choi  \And James Sharpnack \And Tucker Jones }

\aistatsaddress{ UC Davis \And UC Davis \And UC Davis \And Amazon\footnote{Work done prior to joining Amazon} \And UC Davis  } ]

\begin{abstract}
  Strong gravitational lenses allow us to peer into the farthest reaches of space by bending the light from a background object around a massive object in the foreground. 
  Unfortunately, these lenses are extremely rare, and manually finding them in astronomy surveys is difficult and time-consuming.  
  We are thus tasked with finding them in an automated fashion with few if any, known lenses to form positive samples.
  To assist us with training, we can simulate realistic lenses within our survey images to form positive samples.
  Naively training a ResNet model with these simulated lenses results in a poor precision for the desired high recall, because the simulations contain artifacts that are learned by the model.
  In this work, we develop a lens detection method that combines simulation, data augmentation, semi-supervised learning, and GANs to improve this performance by an order of magnitude.
  We perform ablation studies and examine how performance scales with the number of non-lenses and simulated lenses.
  These findings allow researchers to go into a survey mostly ``blind" and still classify strong gravitational lenses with high precision and recall.
 
\end{abstract}

\section{Introduction}

\begin{figure}[h]
    \centering
    \includegraphics[width=\columnwidth]{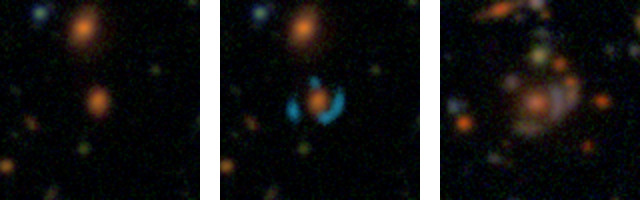}
    \caption{Non-lens (left), simulated lens (middle), real lens (right)}
    \label{fig:non-lens, simulated lens, and real lens}
\end{figure}
 
Massive galaxies can deflect the light from background sources through the effect of gravitational lensing, creating magnified ``arcs'' and multiple images of background galaxies when they are located directly along the line of sight. Such alignments are rare, and these lensing systems are important to astronomers for a range of studies, such as glimpsing into the farthest regions of space where the light of distant objects is ordinarily too faint to detect.  With strong gravitational lenses, this light becomes focused and amplified. Additionally, the lensing information can be used to study the mass distribution in foreground galaxies, notably including the non-baryonic dark matter which comprises most mass in the universe \citep{strongGravitationalLensChallenge}.

A principal challenge is that strong gravitational lenses are incredibly rare. Across the entire sky only of order a thousand such systems are currently known \citep{strongGravitationalLensChallenge}. Previous efforts to find strong gravitational lenses have largely been done manually by individuals visually inspecting images. This is both impractical and expensive. In recent years, various groups have turned to deep learning methods to search for lens systems \citep[e.g.,][]{Colin-CFHTLS,hsc-2018, Milad-HST, colin-des-2019, Huang_DESI-decam_2020, kids-lens-search, panstaars-2020}. These early attempts were rather simplistic as they typically only train and evaluate their models in a supervised fashion either on small numbers of known lenses or by making simulated lenses from their own surveys. Nonetheless, deep learning is proving to be a fruitful and efficient approach. \footnotetext{Work done prior to joining Amazon}

Several surveys are planned for the next decade to observe wide areas of the sky at unprecedented depth and angular resolution (e.g., Rubin Observatory [\citealp{LSST2009}], Euclid [\citealp{euclid2011}] and Roman Space Telescope [\citealp{Roman2015}]). These will enable the detection of orders-of-magnitude more strong lenses than with current data \citep{oguri2010}. The early challenge of analyzing these surveys is that astronomers will not have access to lenses to build their classifiers. In this case, there are two primary options: (1) use lenses found from other surveys and hope the features are effective and transferable, or (2) create simulated lenses based on each survey and train a classifier on those. For (1), the biggest issue is that the transferred performance may vary significantly. This is due to the fact that images from these other surveys are produced with different instruments as well as different preprocessing techniques. Therefore, the samples used for training may be too distributionally dissimilar (i.e.~covariate shift) from their target to be useful. One possibility to ameliorate the effects of covariate shifts is to use CycleGANs \citep{cyclegan} to transform these images to look like the target data distribution. However, it still doesn't solve the issue of the extremely small number of known lenses with heterogeneous imaging. So (2) is the realistic option for producing consistent performance across surveys by simulating lenses directly on the target set. 

Using simulations for training data is quite common in deep learning \citep{Nikolenko2019SyntheticDF}. The problem with option (2) though is that researchers will be creating simulated lenses without a reference point for how they look in their survey. This results in classifiers having good performance when evaluating on held out simulations, but poor performance when classifying real lenses.  This is especially problematic for multi-channel images (see Fig.~\ref{fig:non-lens, simulated lens, and real lens}) since getting the channel information incorrect can lead to an ineffective classifier. Instead of trying to get all the channel information of the arcs correct, one possibility is to simulate lenses on a single channel and build classifiers to detect lenses in this setting \citep{panstaars-2020}. This sidesteps the issue of getting the channel information correct, but this workaround causes us to lose some contextual information about the ``coloring" of the lenses and the surrounding objects, which may actually help the model learn to detect lenses. As a result, we do not explore this option in this paper. We also do not explore using pretrained networks here. Instead, we will focus on a completely self-contained regimen for building classifiers from simulated data. Data augmentation is one way to address this issue of realism without sacrificing this multi-channel information from the image. Secondly, while we can obtain a small sample of non-lensed images to train our classifier, the majority of the survey remains unlabeled, so the use of semi-supervised learning (SSL) algorithms is also a prudent direction to boost the performance of the classifier. By understanding the correct ways to leverage these methods in concert, we can show that you can create highly effective classifiers for detecting lenses even if you only train on potentially ``bad" simulated lenses.

\section{SSL And Unsupervised Learning}


The simplest approach to building a classifier is to use the simulated lenses as our target and train a fully supervised classifier. The limitations of course is that the unlabelled data isn't leveraged and the simulated lens distribution may differ from that of the real lenses.

\subsection{Semi-supervised Learning}

We find that SSL algorithms are another indispensable tool for classifying lenses.
In recent years, the field of deep learning has seen significant progress in the area of semi-supervised learning algorithms \citep{Yang2021ASO, vanEngelen2019ASO}. Instead of covering all of them, we will focus on a narrow collection of state-of-the-art algorithms: Pseudo-label \citep{pseudo-label},  $\Pi$-model \citep{pi_model}, Mean Teacher \citep{mean_teacher}, VAT \citep{mean_teacher}, MixMatch \citep{MixMatch}.

For semi-supervised learning algorithms, there are usually two primary goals: consistency regularization and entropy minimization. 
Some SSL methods (e.g.~consistency regularization) considered here require data augmentation (DA), and we summarize the DA methods used in Table \ref{tab:data_augmentation}.
These methods are chosen specifically with this application in mind.

Consistency regularization is based on the idea that a classifier should output the same predictions even if the image has been augmented. This is usually carried out by appending a regularizing term to the loss that computes the ``distance" between the outputs of the classifier evaluated on two stochastically augmented versions of the same image. Almost all the algorithms we listed above utilize this in some form or another, with the exception of pseudo-label. The set of augmentations is also usually something predefined, which means that the application isn't domain agnostic, and performance will largely depend on the domain-specific augmentations. The exception of course is VAT \citep{VAT}, which generates the augmentations during training instead of being predefined. 

Entropy minimization is based on the idea that the decision boundary of the classifier should lie in low-density regions. Worded another way, if two images $x_1$ and $x_2$ are close in a high-density region then the predictions $y_1$ and $y_2$ should be close as well. Pseudo-label and MixMatch both try to enforce these properties. Pseudo-label does it by assigning pseudo-labels to unlabeled images which are determined by the class with the highest predicted value. MixMatch does this too but less dramatically by sharpening the predicted values to be used as the label instead of hard thresholding the predictions to produce a pseudo-label.


Typically, the SSL setting assumes that the training and test distributions are the same.
However, we will train on simulated images and test on real lenses.
A priori it was unclear if the SSL algorithms would improve the metrics in question.
Furthermore, there is also the question of whether or not SSL algorithms will even improve over baselines tuned with data augmentations since it has been shown in the past that a classifier's performance can often match the state-of-the-art SSL algorithms by choosing the correct data augmentations \citep{oliver2018}.
Nevertheless, we find that SSL algorithms are an indispensable tool in our arsenal.

\vspace{-0.1cm}
\subsection{Unsupervised Learning}

In unsupervised learning, the typical use case is to learn a data distribution. In deep learning, this is typically done by training a GAN \citep{GAN2014, WGAN-2017, wgan_gp}. The outcome of training a GAN is a generator that can produce similar samples from the data distribution it learned from. Those samples are then used for training the classifier. One can think of this as another form of data augmentation. The difference in our case is that the data distribution(i.e.~the simulated lenses) we would use to train our GAN does not come from our target distribution(i.e.~the real lenses). However, we believe that this can still be helpful because, as we mentioned earlier, we have no a priori notion of what a lensed image would look like coming from the DLS survey. And because GANs do not necessarily faithfully reproduce the data distribution it was trained on, this more exotic form of augmentation should nevertheless be beneficial for improving our classifier's ability to generalize to real lenses.

\begin{table*}[h]
    \scriptsize
	\caption{Summary Of Images In Datasets}
	\label{tab:datasets}
	\begin{center}
	\begin{tabular}{cccccc}
		\textbf{Dataset} & \textbf{Non-lenses} & \textbf{Simulated lenses} & \textbf{Real lenses} & \textbf{Unlabeled data} & \textbf{Percentage of lenses}\\
		\hline \\
        TrainingDataPure & 267961 & 259489 & 0 & 0 & -\\
        TrainingV1 & 266301 & 257874 & 0 & 0 & -\\
        TrainingV2 & 7074 & 6929 & 0 & 259248 & -\\
        SimTest & 786 & 773 & 0 & 0 & -\\
        TestV1 & 874 & 0 & 52 & 0 & 5.6156\\
        TestV2 & 874 & 0 & 27 & 0 & 2.9967\\
	\end{tabular}
	\end{center}
\end{table*}

\begin{table*}[h]
\scriptsize
\caption{Augmentations Used During Training}
\label{tab:data_augmentation}
\begin{center}
\begin{tabular}{cc}
\textbf{Name} & \textbf{Description} \\
\hline  \\
RGB-shuffle & Randomly perturb the order of the channels in the images \\
JPEG quality & Randomly apply JPEG compression with quality between 50-100\% \\
Rot90 & Randomly rotate the 
      images by a multiple of 90 degrees    \\ 
Translations & Randomly translate the images by 
              at most 20 pixels in the up, down,
             left and right directions    \\
Horizontal flips & Randomly flips the images
                 across the x-axis \\
Color augmentation &  Randomly perturb the 
                    brightness(-0.1-0.1), saturation(0.9-1.3),
                    hue(0.96-1.00),\\
                   & and gamma(1.23-1.25) of the images \\
\end{tabular}
\end{center}
\end{table*}

\section{Data And Experimental Setup}

The data that will be the focus of our study comes from the Deep Lens Survey \citep[DLS;][]{DLS-Wittman-2002}. 
Due to the paucity of known lenses in this survey, we do not allow any training or validation (model tuning) to be done on real lenses.
Rather they were reserved for the final comparison of a handful of methods attempted.

\subsection{Deep Lens Survey (DLS) and Lens Simulations}\label{sec:dls-sims}
The Deep Lens Survey consists of 5 independent fields of 4 deg$^2$ each, with images taken over $\sim$100 nights using the 4-meter Blanco and Mayall telescopes. The full 20 deg$^2$ area contains $\sim$5 million cataloged galaxies imaged in 4 different astronomical filters (B,V,R,z) which roughly cover the visible spectrum (i.e.~3000-10000\AA). The throughput curve for the filters is published in \citet{DLSthroughput2013} and the data products from the survey are available for public use. For this work, we make use of only the BVR filters as they have the highest SNR. A total of 267,961 galaxies which are likely to act as strong lenses (i.e.~which appear to be massive galaxies at moderate cosmological redshifts) were photometrically selected for this analysis by applying a R band magnitude cut ($17.5<R<22$). Color images are then constructed for these galaxies using HumVI \citep{phil-humvi} with the target galaxies centered in the images.

For any given image from the survey, we create a simulated lens counterpart which we use for training. We assume a background galaxy is present behind the central galaxy, and use the \verb!glafic! \citep{glafic} lens modeling code to trace the background galaxy's light through the foreground lensing potential. We add the resulting simulated lensed arcs to the DLS survey images using HumVI. The values chosen for the background galaxy and the lensing potential used in the simulations do not rely on any physical property of the foreground and background galaxy. Instead, they randomly probe a range of Einstein radii and redshift values, appropriate for the selected target galaxies, yielding a wide variety of lens configurations. Each simulated lens image has exactly one non-lensed image pair. In other words, we do not use the same non-lensed image to create multiple simulated lensed images in different lens configurations. For this work, we limit our simulations to background sources with relatively blue colors, as these are the most common at high redshifts ($z>1$) and the most likely to be detectable in DLS data. Finally, we visually inspected the simulated lenses and removed any images where the central galaxy was significantly brighter than the arc. This resulted in 259,489 simulated lenses.


\subsection{Training Data}
From the non-lenses and the corresponding simulated lenses, we make two training sets: TrainingV1 and TrainingV2. For TrainingV1 we use 266,301 images for non-lenses and 257,874 corresponding simulations as lenses. For TrainingV2 we use the 7,074 human-labeled objects as non-lenses and 6,929 corresponding simulations as lenses. The rest of the 259,248 images serve as unlabelled data. During training, we do a 90-10 split whereby 90 percent of this data is used for training the ResNet model and 10 percent is reserved for validation. We also created a holdout set, SimTest, of 786 images for non-lenses and 773 images for lenses to be explicitly used for testing in the simulated setting. We summarize these details in Table \ref{tab:datasets}.

\subsection{Initial Lens Discovery and Testing Data}
Prior to this work, there were only a few real lenses, with which we might form our test set, known in the entire DLS survey.
Since manually searching the entire survey for lenses (which are very rare) is a laborious and time-consuming task, we use a pilot model to perform an initial search of the survey. The pilot model was built with the convolution neural network ResNet \citep{he2016deep, he2016identity} of 11 layers depth with polar-transformed images as the input. The motivation behind transforming the images to a polar coordinate system is that at lower Einstein radii and galaxy scales, the arcs are approximately symmetric around the image center and a polar transform captures this symmetry as a straight line. The pilot model was an ensemble of 5 ResNet model instances, and each instance was trained on a randomly selected subset of a pilot training dataset which contained 200,000 simulated lenses and 200,000 non-lenses. The entire sample of unlabeled survey images from the DLS survey (279,149 images in total) was scored by the ResNet ensemble. Around 3000 galaxy images (1\% of survey) that had the highest median scores were taken for human inspection by a team of astronomers. 52 were labeled as good lens candidates and form our test set TestV1. 27 out of the 52 were deemed very likely to be strong real lens candidates and form our test set TestV2.

In order to populate our test sets with real non-lenses, the same team of astronomers were asked to label a fraction of images ($\sim$ 3\%) from the survey. Since most of the images are expected to be non-lenses, this is an easy task and does not warrant an ML model. A total of 8734 galaxy images were labelled to be non-lenses from the entire survey and 874 (i.e., 10\%) were randomly chosen out of those to be included as non-lenses for both the TestV1 and TestV2 test sets.
Therefore with the help of the pilot model and human labeling, two test sets: TestV1 (52 Lenses, 874 NonLenses) and TestV2 (27 Lenses, 874 NonLenses) are constructed.

This initial lens discovery was done independently of the experiments in order to prevent data leakage between the real lens test sets and the supervised models trained in the main experiment.
In addition, we set the number of candidate images returned by the pilot model to be much larger than the expected number of lenses in the DLS survey.
As a result, the number of discovered lenses is reasonable given what has been found in other surveys.
With this being the case, we believe that the main source of error in the test set labels is human error in the hand labeling of real lenses.


\subsection{Experimental Setup And Implementations}

We used a standard ResNet-11 architecture for all experiments. The models we use are broadly broken up into 3 groups: supervised, semi-supervised, and GAN+semi-supervised. 
There are four supervised models: SupervisedV1, SupervisedV1+DA, SupervisedV2, and SupervisedV2+DA. SupervisedV1 and SupervisedV2 are trained on TrainingV1 and TrainingV2, respectively, but without data augmentation. SupervisedV1+DA and SupervisedV2+DA are trained on TrainingV1 and TrainingV2, respectively, but with data augmentation.

For the semi-supervised models, we have five: MixMatch, Pseudo-label, Mean Teacher, $\Pi$-model, and VAT. These are all trained on TrainingV2 with data augmentations.
Lastly, we have the GAN+semi-supervised models. The only difference here is that we expand TrainingV2 with 7000 randomly generated lenses from the WGAN we trained on simulated lenses from TrainingV1. The architecture used for the WGAN is nearly the same as the one used in \cite{wgan_gp}, with some modifications. The model selection for the WGAN was based on visual inspection rather than measurements like FID because we aren't trying to perfectly recreate the simulated data distribution. Data augmentation is applied to the GAN+semi-supervised models as well.

We trained each model for 100 epochs using the Adam optimizer with a learning rate of 0.001 and a minibatch size of 32. The only exceptions are the performance scaling experiments where we used 200 and 400 epochs when the number of simulated lenses and non-lenses were 2000 and 1000, respectively.
For all models, we used a 90-10 split for training and validation. We trained four candidates for each model which we used to evaluate the test sets to get an average performance. 
For the pseudo-label models we used $\alpha=0.1$. For $\Pi$-model we used $\lambda_U = 0.1$. For MixMatch we used $T=0.5$, $K=2$, $\alpha=0.75$, $\lambda_U = 0.1$, $\text{ema\_decay}=0.999$. For Mean Teacher we used $\lambda_U=0.1$ and $\text{ema\_decay}=0.999$. For VAT we used $\epsilon=0.001, \epsilon_{\text{adv}} = 0.1$ and $\alpha = 0.01$.

\section{Results}


\begin{table*}[h]
    \scriptsize
	\caption{Average Precision (\%) For TestV1}
	\label{tab:average_TestV1}
	\begin{center}
	\begin{tabular}{ccccccc}
		       \textbf{Model}         & \textbf{Recall 50\%} & \textbf{Recall 60\%} & \textbf{Recall 70\%} & \textbf{Recall 80\%} & \textbf{Recall 90\%} & \textbf{Recall 100\%} \\ 
		  \hline \\
		     SupervisedV1       &           $3.4 \pm 0.10$          &          $3.99 \pm 0.15$           &           $4.35 \pm 0.10$          &           $4.67 \pm 0.07$           &          $5.14 \pm 0.03$           &          $5.62 \pm 0.01$           \\
		     SupervisedV1+DA       &          $18.32 \pm 6.59$         &          $12.70 \pm 5.39$          &           $7.68 \pm 1.78$          &           $5.99 \pm 0.48$           &          $5.71 \pm 0.24$           &          $5.66 \pm 0.03$           \\
		     SupervisedV2       &          $17.93 \pm 9.00$         &          $10.09 \pm 7.62$          &           $8.39 \pm 6.07$          &           $6.85 \pm 3.13$           &          $5.64 \pm 0.85$           &          $5.65 \pm 0.02$           \\
		     SupervisedV2+DA       &         $65.46 \pm 15.00$         &         $55.54 \pm 13.13$          &          $46.45 \pm 14.49$         &          $33.02 \pm 10.26$          &          $22.60 \pm 4.78$          &          $8.13 \pm 2.49$           \\
		       MixMatch         &          $89.67 \pm 6.70$         &         $72.81 \pm 11.95$          &           $54.58 \pm 18.8$         &          $40.77 \pm 11.95$          &         $30.33 \pm 10.06$          &          $12.28 \pm 5.09$          \\
		   Pseudo-label      &         $64.17 \pm 16.67$         &         $54.81 \pm 18.02$          &          $46.20 \pm 23.08$         &          $26.56 \pm 14.91$          &          $10.62 \pm 6.13$          &          $5.75 \pm 0.23$           \\
		     Mean Teacher       &          $81.30 \pm 6.08$         &          $73.75 \pm 4.04$          &           $63.01 \pm 6.61$         &          $43.43 \pm 7.15$           &          $24.43 \pm 2.66$          &          $6.61 \pm 0.99$           \\
		     $\Pi$-Model        &          $75.86 \pm 8.35$         &          $67.71 \pm 9.16$          &          $54.54 \pm 10.09$         &          $43.94 \pm 12.35$          &          $30.74 \pm 3.77$          &          $13.41 \pm 2.33$          \\
		         VAT            &          $68.17 \pm 9.00$         &          $50.47 \pm 9.08$          &           $43.85 \pm 7.39$         &          $33.69 \pm 3.81$           &         $20.72 \pm 11.50$          &           $9.25 \pm 5.7$           \\
			
		   GAN + Supervised     &          $84.95 \pm 8.70$         &          $79.88 \pm 6.30$          &           $69.42 \pm 4.60$         &          $54.35 \pm 4.57$           &          $34.12 \pm 7.47$          &          $8.25 \pm 2.85$           \\
		    GAN + MixMatch      & $\textbf{96.86} \pm \textbf{3.70}$&          $93.93 \pm 4.37$          &  $\textbf{83.90} \pm \textbf{1.92}$&          $63.90 \pm 5.69$           & $\textbf{46.56} \pm \textbf{9.83}$ &          $14.13 \pm 6.53$          \\
		GAN + Pseudo-label  &          $88.3 \pm 7.84$          &          $84.00 \pm 8.18$          &          $78.27 \pm 11.24$         &          $48.16 \pm 7.67$           &         $28.53 \pm 13.21$          &          $6.87 \pm 1.33$           \\
		  GAN + Mean Teacher    &          $94.7 \pm 4.45$          & $\textbf{94.32} \pm \textbf{5.07}$ &           $80.83 \pm 4.92$         &          $67.33 \pm 3.22$           &          $36.95 \pm 6.50$          & $\textbf{15.57} \pm \textbf{6.50}$ \\
		  GAN + $\Pi$-Model     &          $92.89 \pm 1.36$         &          $86.14 \pm 12.7$          &           $79.1 \pm 18.02$         & $\textbf{69.24} \pm \textbf{17.25}$ &         $41.47 \pm 12.87$          &          $15.2 \pm 6.21$           \\
		      GAN + VAT         &         $87.96 \pm 10.23$         &          $75.15 \pm 15.2$          &          $60.51 \pm 12.40$         &          $48.92 \pm 13.48$          &          $24.62 \pm 6.46$          &          $11.15 \pm 8.77$  \\
			
	\end{tabular}
    \end{center}
\end{table*}

\begin{table*}[h]
    \scriptsize
	\caption{Average Precision (\%) For TestV2}
	\label{tab:average_TestV2}
	\begin{center}
	\begin{tabular}{ccccccc}
		       \textbf{Model}         & \textbf{Recall 50\%} & \textbf{Recall 60\%} & \textbf{Recall 70\%} & \textbf{Recall 80\%} & \textbf{Recall 90\%} & \textbf{Recall 100\%} \\ 
			\hline \\
		     SupervisedV1       &          $1.90 \pm 0.09$           &          $2.20 \pm 0.07$           &           $2.39 \pm 0.07$           &           $2.63 \pm 0.06$           &          $2.82 \pm 0.03$           &          $3.01 \pm 0.02$           \\
		     SupervisedV1+DA       &          $8.55 \pm 3.92$           &          $3.81 \pm 0.99$           &           $3.03 \pm 0.15$           &           $2.97 \pm 0.13$           &          $2.92 \pm 0.04$           &          $3.03 \pm 0.03$           \\
		     SupervisedV2       &          $8.55 \pm 3.75$           &          $4.74 \pm 3.41$           &           $4.23 \pm 2.86$           &           $3.69 \pm 1.60$           &          $2.98 \pm 0.17$           &          $3.06 \pm 0.04$           \\
		     SupervisedV2+DA       &         $55.77 \pm 17.43$          &         $44.09 \pm 12.94$          &          $37.24 \pm 12.85$          &          $24.37 \pm 10.50$          &          $10.99 \pm 4.25$          &          $5.79 \pm 3.93$           \\
		       MixMatch         &         $83.54 \pm 11.31$          &         $77.98 \pm 12.13$          &          $51.26 \pm 17.7$           &          $29.44 \pm 13.24$          &          $12.06 \pm 3.17$          &          $6.84 \pm 3.00$           \\
		   Pseudo-label      &         $53.06 \pm 20.53$          &         $47.04 \pm 18.00$          &          $37.44 \pm 23.18$          &           $17.99 \pm 18$            &          $4.18 \pm 2.16$           &          $3.07 \pm 0.13$           \\
		     Mean Teacher       &          $72.35 \pm 6.79$          &          $65.05 \pm 4.41$          &          $60.09 \pm 4.09$           &          $28.96 \pm 13.14$          &          $5.74 \pm 2.81$           &          $3.55 \pm 0.55$           \\
		     $\Pi$-Model        &         $63.33 \pm 10.34$          &          $55.66 \pm 9.53$          &          $44.98 \pm 10.73$          &          $30.48 \pm 8.12$           &          $15.96 \pm 1.09$          &          $8.68 \pm 1.49$           \\
		         VAT            &         $61.93 \pm 10.11$          &         $46.85 \pm 11.96$          &          $40.21 \pm 11.91$          &          $21.37 \pm 4.59$           &          $9.43 \pm 7.97$           &          $7.79 \pm 8.68$           \\
		   GAN + Supervised     &         $80.19 \pm 17.08$          &         $78.03 \pm 12.45$          &          $68.05 \pm 12.96$          &          $40.98 \pm 17.12$          &         $16.33 \pm 8.661$          &          $6.05 \pm 2.69$           \\
		    GAN + MixMatch      & $\textbf{96.88} \pm \textbf{6.25}$ & $\textbf{92.84} \pm \textbf{5.69}$ &          $78.90 \pm 5.65$           &          $39.51 \pm 9.01$           &         $20.54 \pm 11.69$          &          $7.97 \pm 3.93$           \\
		GAN + Pseudo-label  &         $80.58 \pm 10.88$          &         $77.75 \pm 14.94$          &          $70.42 \pm 13.68$          &          $30.36 \pm 3.56$           &          $9.49 \pm 3.91$           &          $3.69 \pm 0.74$           \\
		  GAN + Mean Teacher    &          $89.51 \pm 8.70$          &         $87.99 \pm 12.04$          & $\textbf{81.83} \pm \textbf{14.03}$ &          $46.41 \pm 18.12$          &          $14.61 \pm 5.22$          &          $8.87 \pm 3.92$           \\
		  GAN + $\Pi$-Model     &          $85.62 \pm 2.38$          &          $74.9 \pm 13.99$          &          $69.58 \pm 17.84$          & $\textbf{54.26} \pm \textbf{13.05}$ & $\textbf{30.45} \pm \textbf{8.72}$ & $\textbf{22.27} \pm \textbf{7.71}$ \\
		      GAN + VAT         &         $78.84 \pm 17.21$			 &          $60.40 \pm 4.77$		  &          $44.46 \pm 10.12$			&          $27.29 \pm 7.02$			  &          $11.99 \pm 5.15$		   &          $8.63 \pm 5.71$
	\end{tabular}
    \end{center}
\end{table*}

\subsection{Data Augmentation Significantly Improves Performance}

From our results, we see that data augmentations have a significant effect on a classifier's performance. This improvement can be seen from SupervisedV2 and SupervisedV2+DA where we see a  5-10 times performance uplift when recall is below 80\% in Table \ref{tab:average_TestV1} and Table \ref{tab:average_TestV2}. This is in line with expectations since we assumed that simulated lenses would look different from their real counterparts. The good news is that data augmentation can compensate for this, and it substantially closes the gap so that the classifier is usable. Higher recall levels are more challenging however and we see this reflected in those tables as well. For recall above 80\%, we see that data augmentation can no longer compensate, which suggests that the simulated data in the training sets were unable to capture the relevant details for those remaining cases. 

\subsection{SSL Algorithms Lead The Charts}

When we look at average precision, we see that the most performant algorithms are all SSL algorithms. Adding GAN-generated lenses to the training set also offers a meaningful boost to performance. From Table \ref{tab:average_TestV1} and Table \ref{tab:average_TestV2} we see that GAN+MixMatch is first or second in terms of precision at all recall levels. This model is also a substantial improvement over the baseline supervised model, SupervisedV2+DA, by nearly doubling the performance in precision at all recall levels.


Let us summarize some key findings from Table \ref{tab:average_TestV1} and Table \ref{tab:average_TestV2}. The first thing to notice is that training with more non-lenses seems to hurt performance. This seemingly counterintuitive result can be seen in the performance gap between SupervisedV1+DA and SupervisedV2+DA. Note that this decrease in performance only really happens when the additional non-lenses are used in a supervised fashion. On the other hand, when we treat the additional non-lenses as unlabeled and use them in the SSL algorithms we see a boost in performance. This is most likely due to the fact that supervised losses tend to ``push" the decision boundary for the non-lenses into regions where true lenses reside; whereas the unsupervised losses tend to regulate and refine the preexisting boundary defined by the labeled non-lenses, causing a ``pull" or retraction of the decision boundary. Furthermore, this pull effect moreso comes from consistency regularization rather than entropy minimization since we see roughly the same performance between SupervisedV2+DA and Pseudo-label, which does pure entropy minimization. This also suggests that using more simulated or GAN-generated lenses should be beneficial because we want to push the decision boundary into the regions where true lenses reside. Lastly, the improvement in performance from using GAN generated samples also suggests possibly that asymmetry is important between the non-lenses and simulated lenses. By this, we mean that the source images for generating the simulated lenses should not also be used in training as non-lenses.
We explain and justify these findings in more detail in the following sections.

\subsection{More Non-lenses In The Training Set Eventually Decreases Performance}

\begin{figure}[ht]
\centering

\subfloat[]{{\includegraphics[width=0.5\columnwidth]{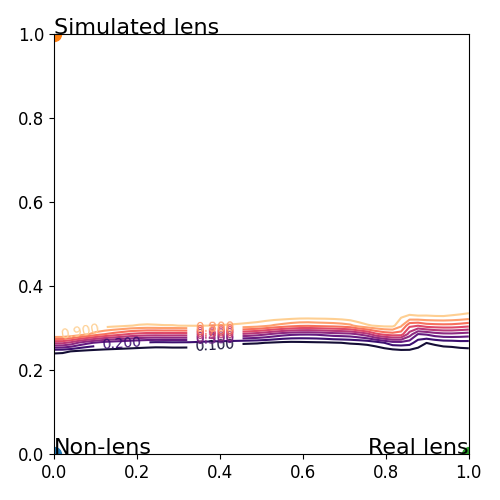}}}%
\subfloat[]{{\includegraphics[width=0.5\columnwidth]{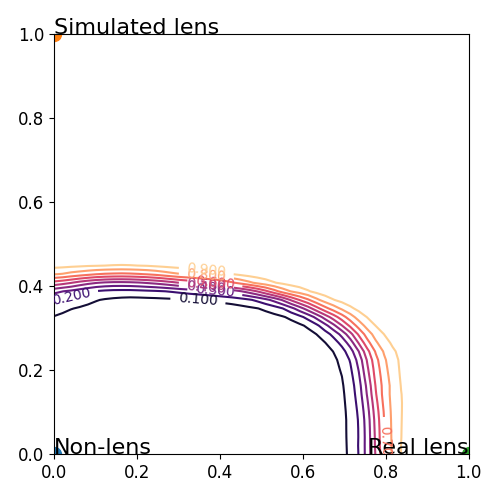}}}%
\quad
\subfloat[]{{\includegraphics[width=0.5\columnwidth]{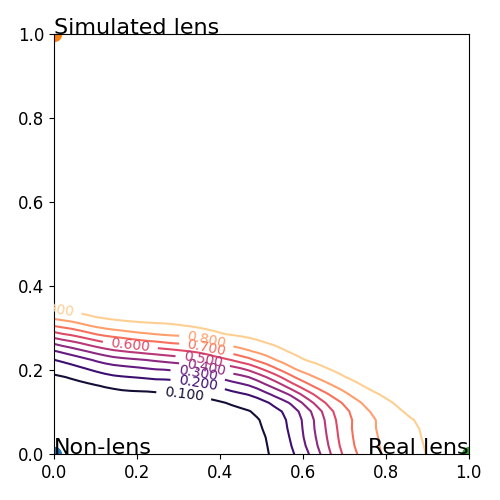}}}%

\caption{Linear interpolation between randomly sampled non-lens, simulated lens, and real lens for methods: SupervisedV1+DA (top left), SupervisedV2+DA (top right), GAN+MixMatch (bottom).}
\label{fig: contour}
\end{figure}

Recall that TrainingV1 is the larger training set that consists of nearly the whole survey and the corresponding simulated lenses. TrainingV2 on the other hand is the smaller training set made from human-labeled non-lenses and the corresponding simulated lenses. As we mentioned before, there is a noticeable performance gap between models trained on TrainingV1 versus TrainingV2. This can be seen from the difference in precision between SupervisedV1+DA and SupervisedV2+DA in Table \ref{tab:average_TestV1} and Table \ref{tab:average_TestV2}. This performance gap isn't due to overfitting since validation and testing accuracy on simulated data was nearly perfect. Instead, the drop in performance seems to be caused by the increasing likelihood of non-lenses, which look like lenses, being present in the data that pushes the non-lens decision boundary further into the regions where real lenses reside. To see this we randomly sampled images of non-lenses($z_{nl}$), simulated lenses($z_{sl}$) and real lenses($z_{rl}$) and interpolated between them($z$) to see how the decision boundary changes between these points. 

\begin{equation}
    \label{eq: interpolated contour equation}
    z=z_{nl}(1-x-y) +z_{sl}(y) + z_{rl}(x) 
\end{equation}

Contour plots of the predicted likelihood that the image is a lens are given in Figure \ref{fig: contour}, where the $x$-axis and $y$-axis correspond to the $x$ and $y$ values in the interpolation formula above. Notice how the contour lines are pushed further away from the real lens in SupervisedV1+DA than SupervisedV2+DA. We also notice that the contour lines are bundled in a much tighter neighborhood for SupervisedV1+DA than SupervisedV2+DA. The effect of tighter neighborhoods is poorer generalization because it leads to less granularity for thresholding. This increases the likelihood of false positives being selected since the regions intersect more highly with the regions containing real lenses. If we look at GAN+MixMatch, we see that we can overcome these limitations by adding GAN-generated lenses to the training set. Notice also that the decision boundary gets pushed further into the region where real lenses reside allowing a threshold on the output to reliably distinguish between real lenses and non-lenses.


To further confirm our assertions we ran another set of experiments where we used TrainingV1 but varied the number of non-lenses in the training set. We train a supervised model on these different training sets. The results can be seen in Table \ref{tab:performance scaling non-lenses}. Overall, we see that while increasing the number of non-lenses increases precision initially, it will eventually peak and from there it dramatically falls in value until it plateaus.



\begin{table*}[h]
    \scriptsize
    \caption{Performance Scaling Of Number Of Non-lenses In TrainingV1 On TestV1 Precision(\%)}
    \label{tab:performance scaling non-lenses}
    \begin{center}
    \begin{tabular}{ccccccc}
        \textbf{Number of non-lenses} & \textbf{Recall 50\%} & \textbf{Recall 60\%} & \textbf{Recall 70\%} & \textbf{Recall 80\%} & \textbf{Recall 90\%} & \textbf{Recall 100\%} \\ \hline \\

        1000                 & 16.61      & 12.5       & 9.38       & 8.46       & 6.66       & 5.76        \\
        2000                 & 44.72      & 23.71      & 21.29      & 12.36      & 5.73       & 5.62        \\
        4000                 & 33.55      & 33.34      & 24.79      & 13.8       & 8.89       & 5.7         \\
        8000                 & 56.56      & 52.46      & 38.96      & 30.34      & 13.9       & 5.96        \\
        16000                & 50         & 42.39      & 37.95      & 34.15      & 14.35      & 8.14        \\
        32000                & 49.07      & 35.56      & 33.49      & 26.38      & 17.4       & 8.33        \\
        64000                & 11.71      & 9.69       & 9.09       & 6.86       & 5.75       & 5.62        \\
        128000               & 13.17      & 8.14       & 6.04       & 5.48       & 5.34       & 5.62        \\
        256000               & 17.10      & 10.52      & 6.83       & 5.12       & 5.33       & 5.66       

    \end{tabular}
    \end{center}
\end{table*}

\subsection{Performance Breakdown By Object Type}

\begin{figure}[h]
    \centering
    \includegraphics[width=0.75\columnwidth]{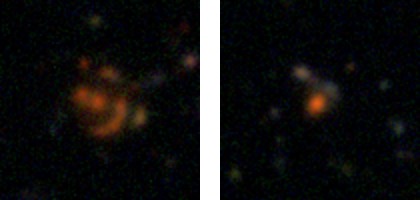}
    \caption{Frequently misclassified lenses}
    \label{fig:misclassified lenses}
\end{figure}
We also looked into whether the performance of the models was consistent based on the properties of the central galaxy in each image. From our analysis, there wasn't any indication that the particular galaxy type affected the model's ability to correctly classify lenses. This is reassuring since the lens simulations used in our training data (see \ref{sec:dls-sims}) are not dependent on any physical property of the central galaxy. The performance of models also remains consistent across different fields of the survey. Instead, the main obstacle seems to just be the coloring of the arc and the brightness of the arc itself. As we can see in Figure \ref{fig:misclassified lenses}, orange arcs are misclassified as non-lenses simply because our training set doesn't account for them. Data augmentations by themselves cannot compensate for this because there is no way to turn the red-orange object in the middle and the blue arc to simultaneously become red-orange in color. There are also faint arcs that we can't properly capture because the arcs used in our simulations are comparatively brighter.

\subsection{Data Augmentation Ablation Study}

\begin{table*}
    \scriptsize
    \caption{Ablation performance for $80\%$ Recall}
    \label{tab:ablation80}
    \begin{center}
    \begin{tabular}{ccccc}
        \textbf{Augmentation removed} & \textbf{TestV1 Precision(\%)} & \textbf{TestV2 Precision(\%)} & \textbf{TestV1 difference(\%)} & \textbf{TestV2 difference(\%)}\\ 
        \hline \\
        - & $54.35 \pm 4.57$ & $40.98 \pm 17.12$ & - & -\\
        GAN & $33.02 \pm 10.26$ & $24.37 \pm 10.50$ & \color{red} -21.33 & \color{red} -16.61\\
        RGB shuffle & $15.34 \pm 7.59$ & $11.75 \pm 4.90$ & \color{red} -39.01 & \color{red} -29.23\\
        JPEG quality & $21.33 \pm 8.29$ & $15.25 \pm 3.41$ & \color{red} -33.02 & \color{red} -25.73\\
        Rot90 & $62.63 \pm 12.43$ & $52.2 \pm 13.65$ & \color{ForestGreen} +8.28 & \color{ForestGreen} +11.22\\
        Translations & $52.17 \pm 12.92$ & $35.09 \pm 16.64$ & \color{red} -2.18 & \color{red} -5.89\\
        Horizontal flips & $60.54 \pm 9.78$ & $36.46 \pm 9.03$ & \color{ForestGreen} +6.19 & \color{red} -4.52\\
        Color augmentation & $39.47 \pm 13.24$ & $33.84 \pm 10.92$ & \color{red} 14.88 & \color{red} -7.14\\

    \end{tabular}
    \end{center}
\end{table*}

 The ablation study was performed with GAN+Supervised as our model. As we can see from Table \ref{tab:ablation80} (see Tables 7-12 in the Supplementary section for more recall values), the most impactful augmentations were RGB shuffle, JPEG quality and GAN. Nearly, 40\% of the precision we are seeing comes directly from RGB shuffling. The reason RGB shuffle is one of the most effective augmentations is that we don't know a priori how real lenses would look in our survey. As a result, the simulated lens we produce will never perfectly align with the channel intensities of the arcs from real lenses. By shuffling the channels we allow the model to be less sensitive on the coloration of the arc and more on the structural properties of the arc in relation to the surrounding objects. This doesn't mean that it's not important to create realistically colored arcs for the simulations, but rather we can compensate for it when it is significantly off from the coloration of arcs around real lenses. JPEG quality and GAN also were significant contributors to the overall performance of the model. 





\subsection{Spectroscopic Lens Confirmation}

\begin{figure}[h]
    \centering
    \includegraphics[width=0.75\columnwidth]{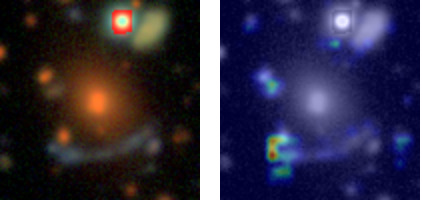}
    \caption{Spectroscopically confirmed lens system DLS212072337(left), Grad-CAM++ overlay from GAN+MixMatch(p-value: 1.00, right), Red-Green-Blue in overlay indicate High-Mid-Low importance of the feature to the model when making the prediction.}
    \label{fig:confirmed lens}
\end{figure}

To confirm the lensing nature of these systems, spectroscopic redshifts are required to establish that the candidate deflector is indeed a massive galaxy, and that the arc is located at a greater distance (higher redshift) than the deflector. We obtained spectra of a high confidence lens found from this work (DLS212072337; see Figure \ref{fig:confirmed lens}) using the NIRES instrument \citep{NIRES2004} at W.M. Keck Observatory on the night of 31 March 2021, and measured a secure redshift for the arc of $z=1.81$ via [O~{\sc iii}]~$\lambda\lambda$4959,5007 and H$\alpha$ emission lines (Tran et al., in prep). The central deflector is known to be a massive galaxy at a lower redshift of $z=0.43$ from BOSS \citep{SDSSBOSS}, thus confirming this as a true lens system. While this is only a single object, these results demonstrate the feasibility of characterizing a larger sample identified from this work.

\section{Discussion}

We have shown in this paper that combining data augmentation, SSL algorithms, and GAN-generated lenses significantly improves the lens classifier performance. In terms of data augmentation, RGB-shuffle produced a significant boost to the lens classifier's performance.
Of the SSL algorithms, we found that consistency regularization is the dominant factor as to why they outperform supervised baselines. 
Figure \ref{fig: contour} shows us that consistency regularization more definitively separates the non-lenses from the lenses.
Our findings indicate that one must take care when balancing the number of non-lenses and simulated lenses in the training data, as increasing one does not necessarily translate to better performance (see Table \ref{tab:performance scaling non-lenses}).

In addition to improving the coverage of non-lenses, we also realized that a greater diversity of simulated lenses was needed.
Data augmentation is unable to compensate for cases where arcs are not blue or when arcs are relatively faint in real lenses. 
One direction worth exploring is if there are any beneficial augmentations that can be applied to the arcs themselves.
Since simulated lenses consist of superimposing an image of an arc onto a non-lens image these augmentations can also be generated on the fly. The effect of additional data augmentations specific to the survey (e.g., varying the Point Spread Function) might also be worth exploring. Another possible improvement would be to create multiple different classes of simulated lenses based on the binned color and brightness of the arcs, turning this binary classification problem into a multiclass classification problem. Breaking up the lens class into multiple subclasses may also offer improvements in performance as well \citep{hoffman2001, luo2008}. 

An additional approach that we didn't explore in this paper is to take the labeled data and combine it with these various simulated lenses and learn representations through self-supervised learning\citep{simclr,SwAV, bootstrapyourownlatent}. Recently, other researchers have started looking into these approaches \citep{hayat2021, george2021}.
The latter preprint is applied to lens discovery in the much larger DESI survey, and simulations are not used in training. Another similar approach by \citep{Cheng2020} utilizes a convolutional autoencoder to construct an embedding of the simulated strong lens images and fit a Gaussian mixture model to extract the clusters in the embedding, which corresponds to various visual features of the galaxies and lenses. 
The results are not directly comparable to our own due to differences in the experimental setup. 

In this work, we use a two-stage process where a pilot model is used to obtain the test set and then use this to select the best out of a handful of models to obtain better candidate lenses.
In practice, one may want to try out several variations of our top-performing methods without the two-stage approach.
One way to achieve this is to ``recommend'' to the astronomer potential lenses in a streaming fashion.
We can then associate a reward to each newly discovered lens and use a multi-armed bandit algorithm to perform online model selection.

In summary, we have isolated several key ingredients that are essential to training a lens classifier using only simulations in an astronomical survey.
This method is able to improve on existing models by an order of magnitude, and it has already led to the discovery and confirmation of a novel gravitational lens system.


\bibliography{references.bib}

\clearpage
\appendix

\thispagestyle{empty}

\onecolumn
\makesupplementtitle

\section{Additional Ablation Results}



\begin{table*}[hbt!]
    \scriptsize
    \caption{Ablation performance for $50\%$ Recall}
    \label{tab:ablation50}
    \begin{center}
    \begin{tabular}{ccccc}
        {Augmentation removed} & {TestV1 Precision(\%)} & {TestV2 Precision(\%)} & {TestV1 difference(\%)} & {TestV2 difference(\%)}\\ 
        \hline \\
        -                  & $84.95 \pm 8.70$ & $80.19 \pm 17.08$ & - & - \\
        GAN                & $65.46 \pm 15.00$ & $55.77 \pm 17.43$ & \color{red} -19.49 & \color{red} -24.42 \\
        RGB shuffle        & $44.8 \pm 24.32$ & $35.45 \pm 21.75$ & \color{red} -40.15 & \color{red} -44.74\\
        JPEG quality       & $65.30 \pm 13.84$ & $56.84 \pm 14.06$ & \color{red}-19.65 & \color{red} -23.35\\
        Rot90              & $91.81 \pm 4.41$ & $89.96 \pm 6.33$ & \color{green}+6.86 & \color{green} +9.77\\
        Translations       & $89.47 \pm 10.34$ & $84.59 \pm 13.04$ & \color{green}+4.52 & \color{green} +4.4\\
        Horizontal flips   & $84.63 \pm 10.85$ & $75.74 \pm 13.96$ & \color{red}-0.32 & \color{red} -4.45\\
        Color augmentation & $71.88 \pm 17.35$ & $68.23 \pm 15.2$ & \color{red}-13.07 & \color{red} -11.96\\

    \end{tabular}
    \end{center}
\end{table*}


\begin{table*}[hbt!]
    \scriptsize
    \caption{Ablation performance for $60\%$ Recall}
    \label{tab:ablation60}
    \begin{center}
    \begin{tabular}{ccccc}
        {Augmentation removed} & {TestV1 Precision(\%)} & {TestV2 Precision(\%)} & {TestV1 difference(\%)} & {TestV2 difference(\%)}\\ 
        \hline \\
        - & $79.88 \pm 6.30$ & $78.03 \pm 12.45$ & - & -\\
        GAN & $55.54 \pm 13.13$ & $44.09 \pm 12.94$ & \color{red} - 24.34 & \color{red} -33.94\\
        RGB shuffle & $34.25 \pm 24.49$ & $26.14 \pm 15.9$ & \color{red} - 45.63 & \color{red} -51.89\\
        JPEG quality & $52.75 \pm 24.68$ & $51.69 \pm 17.91$ & \color{red} - 27.13 & \color{red} -26.34\\
        Rot90 & $84.94 \pm 7.04$ & $78.99 \pm 5.23$ & \color{green} +5.06 & \color{green} + 0.96\\
        Translations & $74.95 \pm 16.11$ & $72.26 \pm 24.65$ & \color{red} -4.93 & \color{red} -5.77\\
        Horizontal flips & $71.15 \pm 11.46$ & $60.64 \pm 12.27$ & \color{red} -8.73 & \color{red} -17.39\\
        Color augmentation & $63.24 \pm 15.67$ & $50.91 \pm 15.53$ & \color{red} -16.64 & \color{red} -27.12\\

    \end{tabular}
    \end{center}
\end{table*}


\begin{table*}[hbt!]
    \scriptsize
    \caption{Ablation performance for $70\%$ Recall}
    \label{tab:ablation70}
    \begin{center}
    \begin{tabular}{ccccc}
        {Augmentation removed} & {TestV1 Precision(\%)} & {TestV2 Precision(\%)} & {TestV1 difference(\%)} & {TestV2 difference(\%)}\\ 
        \hline \\
        - & $69.42 \pm 4.60$ & $68.05 \pm 12.96$ & - & -\\
        GAN & $46.45 \pm 14.49$ & $37.24 \pm 12.85$ & \color{red} -22.97 & \color{red} -30.81\\
        RGB shuffle & $26.68 \pm 17.28$ & $18.75 \pm 11.78$ & \color{red} -42.74 & \color{red} -49.3\\
        JPEG quality & $40.38 \pm 18.64$ & $40.52 \pm 21.74$ & \color{red} - 29.04 & \color{red} -27.53\\
        Rot90 & $79.97 \pm 12.15$ & $73.69 \pm 7.11$ & \color{green} +10.55 & \color{green} +5.64\\
        Translations & $67.69 \pm 12.67$ & $59.31 \pm 22.34$ & \color{red} -1.73 & \color{red}-8.74\\
        Horizontal flips & $67.36 \pm 11.95$ & $55.94 \pm 14.71$ & \color{red}-2.06 & \color{red}-12.11\\
        Color augmentation & $52.79 \pm 14.92$ & $45.93 \pm 13.97$ & \color{red}-16.63 & \color{red} -22.12\\

    \end{tabular}
    \end{center}
\end{table*}

\begin{table*}[hbt!]
    \scriptsize
    \caption{Ablation performance for $80\%$ Recall}
    \label{tab:ablation80_duplicate}
    \begin{center}
    \begin{tabular}{ccccc}
        {Augmentation removed} & {TestV1 Precision(\%)} & {TestV2 Precision(\%)} & {TestV1 difference(\%)} & {TestV2 difference(\%)}\\ 
        \hline \\
        - & $54.35 \pm 4.57$ & $40.98 \pm 17.12$ & - & -\\
        GAN & $33.02 \pm 10.26$ & $24.37 \pm 10.50$ & \color{red} -21.33 & \color{red} -16.61\\
        RGB shuffle & $15.34 \pm 7.59$ & $11.75 \pm 4.90$ & \color{red} -39.01 & \color{red} -29.23\\
        JPEG quality & $21.33 \pm 8.29$ & $15.25 \pm 3.41$ & \color{red} -33.02 & \color{red} -25.73\\
        Rot90 & $62.63 \pm 12.43$ & $52.2 \pm 13.65$ & \color{green} +8.28 & \color{green} +11.22\\
        Translations & $52.17 \pm 12.92$ & $35.09 \pm 16.64$ & \color{red} -2.18 & \color{red} -5.89\\
        Horizontal flips & $60.54 \pm 9.78$ & $36.46 \pm 9.03$ & \color{green} +6.19 & \color{red} -4.52\\
        Color augmentation & $39.47 \pm 13.24$ & $33.84 \pm 10.92$ & \color{red} 14.88 & \color{red} -7.14\\

    \end{tabular}
    \end{center}
\end{table*}

\clearpage

\begin{table*}[hbt!]
    \scriptsize
    \caption{Ablation performance for $90\%$ Recall}
    \label{tab:ablation90}
    \begin{center}
    \begin{tabular}{ccccc}
        {Augmentation removed} & {TestV1 Precision(\%)} & {TestV2 Precision(\%)} & {TestV1 difference(\%)} & {TestV2 difference(\%)}\\ 
        \hline \\
        - & $34.12 \pm 7.47$ & $16.33 \pm 8.661$ & - & -\\
        GAN & $22.60 \pm 4.78$ & $10.99 \pm 4.25$ & \color{red} -11.52 & \color{red} -5.34\\
        RGB shuffle & $10.37 \pm 4.68$ & $5.88 \pm 2.61$ & \color{red} -23.75 & \color{red} -10.45\\
        JPEG quality & $15.15 \pm 6.37$ & $6.18 \pm 2.41$ & \color{red} -18.97 & \color{red} -10.15\\
        Rot90 & $40.72 \pm 12.74$ & $21.79 \pm 3.08$ & \color{green} +6.6 & \color{green} +5.46\\
        Translations & $32.88 \pm 3.35$ & $16.14 \pm 6.72$ & \color{red} -1.24 & \color{red} -0.19\\
        Horizontal flips & $30.81 \pm 20.44$ & $14.72 \pm 6.59$ & \color{red} -3.31 & \color{red} -1.61\\
        Color augmentation & $23.86 \pm 8.13$ & $14.25 \pm 8.45$ & \color{red} -10.26 & \color{red} -2.08\\
    \end{tabular}
    \end{center}
\end{table*}

\begin{table*}[hbt!]
    \scriptsize
    \caption{Ablation performance for $100\%$ Recall}
    \label{tab:ablation100}
    \begin{center}
    \begin{tabular}{ccccc}
        {Augmentation removed} & {TestV1 Precision(\%)} & {TestV2 Precision(\%)} & {TestV1 difference(\%)} & {TestV2 difference(\%)}\\ 
        \hline \\
        - & $8.25 \pm 2.85$ & $6.05 \pm 2.69$  & - & -\\
        GAN & $8.13 \pm 2.49$ & $5.79 \pm 3.93$ & \color{red} -0.12 & \color{red} -0.26\\
        RGB shuffle & $6.43 \pm 0.97$ & $3.84 \pm 1.02$ & \color{red} -1.82 & \color{red} -2.21\\
        JPEG quality & $5.88 \pm 0.21$ & $4.14 \pm 2.09$ & \color{red} -2.37 & \color{red} -1.91\\
        Rot90 & $14.76 \pm 8.42$ & $13.00 \pm 5.16$ & \color{green} +6.51 & \color{green} +6.95\\
        Translations & $10.83 \pm 2.46$ & $7.37 \pm 3.66$ & \color{green} +2.58 & \color{green} +1.32\\
        Horizontal flips & $15.74 \pm 3.06$ & $8.86 \pm 1.84$ & \color{green} +7.49 & \color{green} +2.81\\
        Color augmentation & $11.42 \pm 7.25$ & $6.84 \pm 4.12$ & \color{green} +3.17 & \color{green} +0.79\\
    \end{tabular}
    \end{center}
\end{table*}

\clearpage

\section{Performance Breakdown By Object Type}

\begin{figure}[hbt!]
    \centering
    \includegraphics[width=0.8\textwidth]{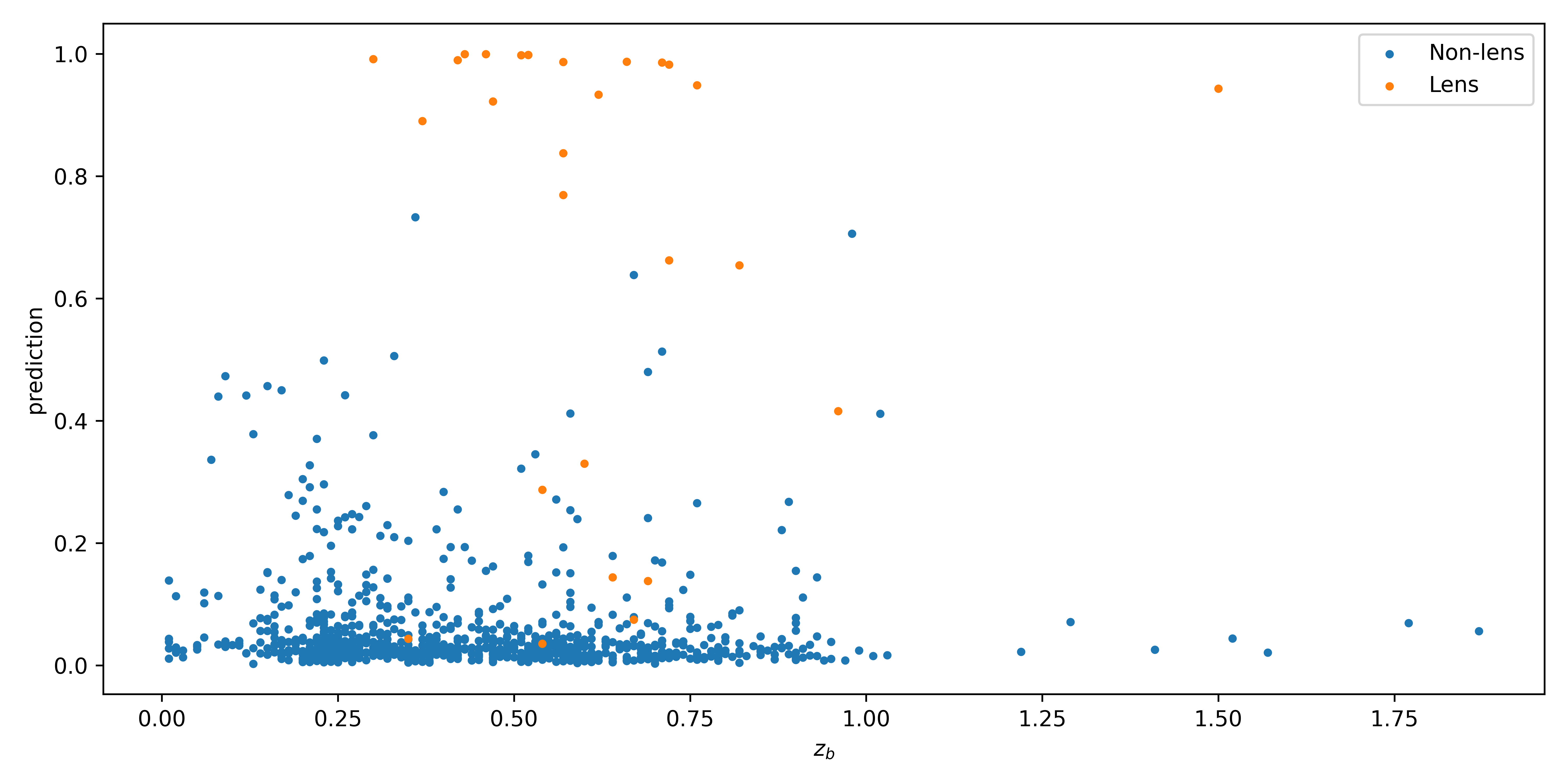}
    \caption{GAN+MixMatch plot of predictions versus the photometric redshift($z_b$) for images in TestV2}
    \label{fig: GAN+MixMatch predictions versus redshift}
\end{figure}

\begin{figure}[hbt!]
    \centering
    \includegraphics[width=\textwidth]{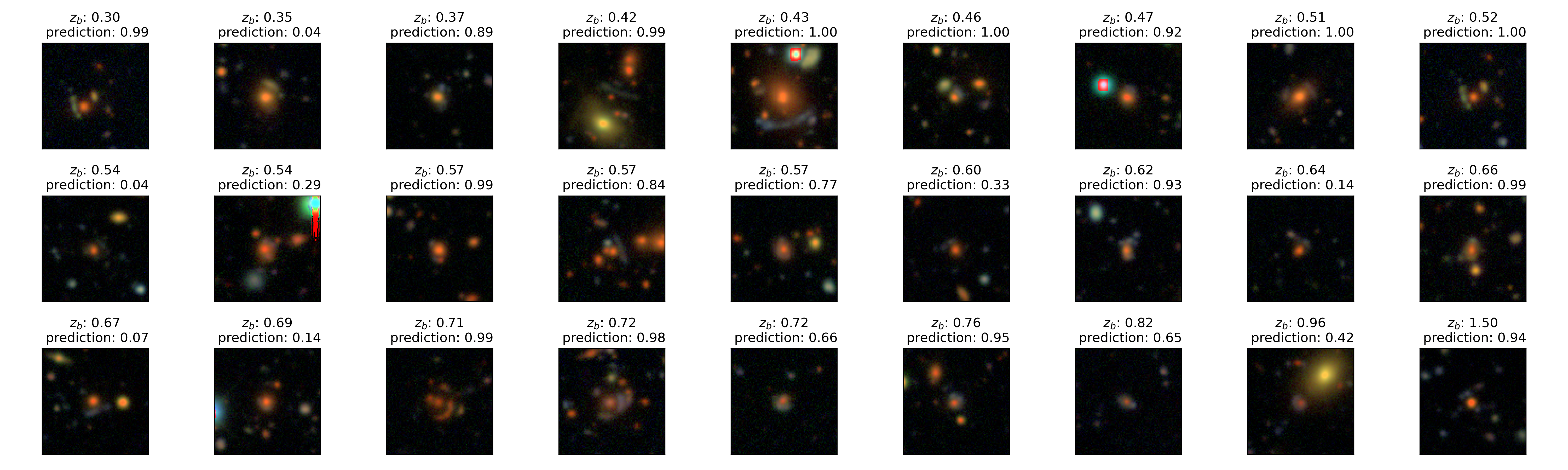}
    \caption{Lenses from TestV2 with prediction from GAN+MixMatch}
    \label{fig:GAN+MixMatch real lenses}
\end{figure}

\begin{figure}[hbt!]
    \centering
    \includegraphics[width=\textwidth]{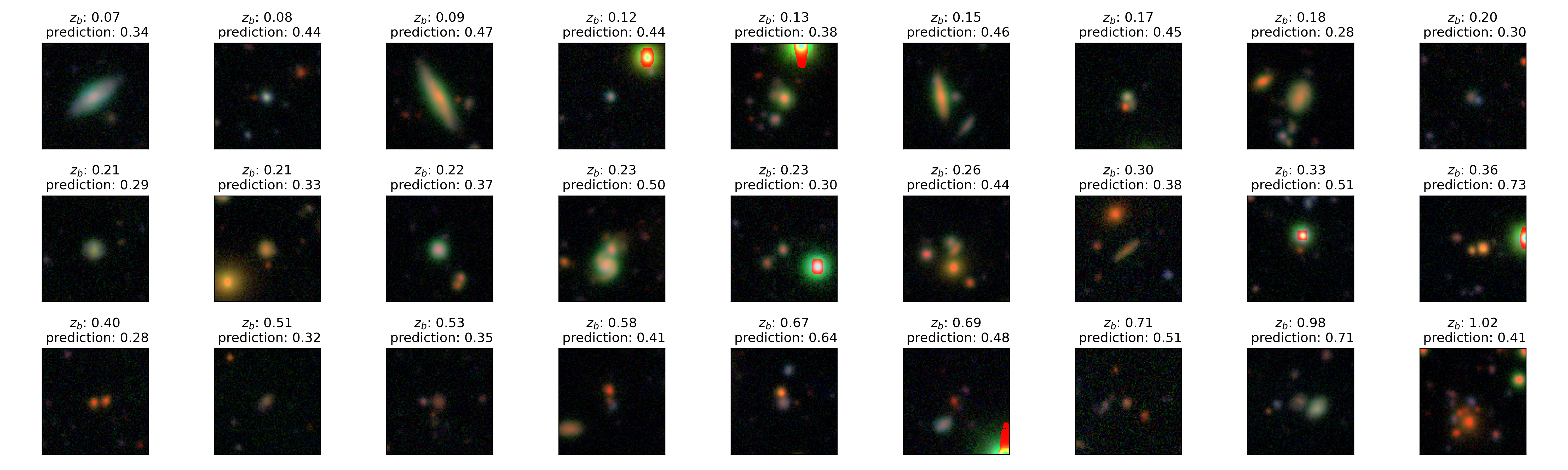}
    \caption{Non-lenses from TestV2 with the highest predictions from GAN+MixMatch}
    \label{fig:GAN+MixMatch non-lenses}
\end{figure}

\clearpage
\section{FixMatch Performance On TestV1 And TestV2}
\begin{table*}[h]
    \scriptsize
	\caption{Average Precision (\%) For TestV1}
	\label{tab:FixMatch average_TestV1}
	\begin{center}
	\begin{tabular}{ccccccc}
		       \textbf{Model}         & \textbf{Recall 50\%} & \textbf{Recall 60\%} & \textbf{Recall 70\%} & \textbf{Recall 80\%} & \textbf{Recall 90\%} & \textbf{Recall 100\%} \\ 
		  \hline \\
		        FixMatch & 22.48 $\pm$ 2.17                & 19.83 $\pm$ 4.04                & 17.42 $\pm$ 5.26                & 13.89 $\pm$ 5.32                & 9.29 $\pm$ 5.58                 & 6.98 $\pm$ 2.63                  \\
		  GAN + FixMatch & 65.88 $\pm$ 14.34               & 46.21 $\pm$ 17.24               & 32.57 $\pm$ 9.00                & 20.95 $\pm$ 5.68                & 15.94 $\pm$ 6.27                & 6.47 $\pm$ 0.50                  \\
			GAN + FixMatch + SA & 96.49 $\pm$ 4.88                & 90.48 $\pm$ 8.61                & 78.40 $\pm$ 10.43               & 64.39 $\pm$ 9.06                & 45.68 $\pm$ 14.84               & 26.48 $\pm$ 9.91  
			
	\end{tabular}
    \end{center}
\end{table*}

\begin{table*}[h]
    \scriptsize
	\caption{Average Precision (\%) For TestV2}
	\label{tab:FixMatch average_TestV2}
	\begin{center}
	\begin{tabular}{ccccccc}
		       \textbf{Model}         & \textbf{Recall 50\%} & \textbf{Recall 60\%} & \textbf{Recall 70\%} & \textbf{Recall 80\%} & \textbf{Recall 90\%} & \textbf{Recall 100\%} \\ 
			\hline \\
		        FixMatch & 13.49 $\pm$ 3.28      & 11.40 $\pm$ 2.48          &  10.62 $\pm$ 3.76            & 9.78 $\pm$ 3.74      & 5.89 $\pm$ 2.43  &  4.44 $\pm$  1.11             \\
		  GAN + FixMatch &  46.31 $\pm$ 20.24                & 31.10 $\pm$ 13.88             & 20.23 $\pm$ 6.83                & 10.81 $\pm$ 2.43               &  6.70 $\pm$ 2.69               & 3.47 $\pm$ 0.28                 \\
			GAN + FixMatch + SA & 91.50 $\pm$ 6.66                 & 86.37 $\pm$ 12.11                & 81.46 $\pm$ 13.67             & 58.92 $\pm$ 16.61                & 33.73 $\pm$ 15.91               & 24.81 $\pm$ 13.82
			
	\end{tabular}
    \end{center}
\end{table*}

FixMatch\citep{fixmatch} is another SSL algorithm that we considered post-hoc. Although it wasn't available at the time of the original study, we are including additional results here to be thorough. The FixMatch algorithm stipulates the use of only weak augmentations in the supervised loss which means it can't deal with generalizing from ``bad" simulated data very well. This can be seen from the results in Tables \ref{tab:FixMatch average_TestV1}-\ref{tab:FixMatch average_TestV2}. By allowing strong augmentations(SA) such as those in Table \ref{tab:data_augmentation}, we are able to salvage this method and make it perform comparable to the other SSL algorithms we have looked at previously.

\end{document}